\documentclass{aastex631}

\hypersetup{linkcolor=red,citecolor=blue,filecolor=cyan,urlcolor=magenta}

\def\lessim{\mathrel{\hbox{\rlap{\hbox{\lower4pt\hbox{$\sim$}}}\hbox{$<$}}}}
\def\grtsim{\mathrel{\hbox{\rlap{\hbox{\lower4pt\hbox{$\sim$}}}\hbox{$>$}}}}

\usepackage{CJK}

\shorttitle{M31N 2017-01e}
\shortauthors{Shafter et al.}

\graphicspath{{./}{figures/}}

\begin{document}
\begin{CJK*}{UTF8}{gbsn}
\title{M31N 2017-01e: Discovery of a Previous Eruption in this Enigmatic Recurrent Nova}

\correspondingauthor{A. W. Shafter}
\email{ashafter@sdsu.edu}

\author[0000-0002-1276-1486]{Allen W. Shafter}
\affiliation{Department of Astronomy, San Diego State University, San Diego, CA 92182, USA}

\author[0000-0002-8482-8993]{Kenta Taguchi}
\affiliation{Department of Astronomy, Kyoto University, Kitashirakawa-Oiwake-cho, Sakyo-ku, Kyoto 606-8502, Japan}

\author[0000-0002-2770-3481]{Jingyuan Zhao (赵经远)}
\affiliation{Xingming Observatory, Mount Nanshan, Xinjiang, China}
\affiliation{Shandong Astronomical Society, Weihai, Shandong 264209, China}

\author[0000-0002-0835-225X]{Kamil Hornoch}
\affiliation{Astronomical Institute of the Czech Academy of Sciences, Fri\v{c}ova 298, CZ-251 65 Ond\v{r}ejov, Czech Republic}

\begin{abstract}

We report the discovery of a previously unknown eruption of the recurrent nova M31N 2017-01e that took place on 11 January 2012. The earlier eruption was detected by Pan-STARRS and occurred 1847 days (5.06~yr) prior to the eruption on 31 January 2017 (M31N 2017-01e). The nova has now been seen to have had a total of four recorded eruptions (M31N 2012-01c, 2017-01e, 2019-09d, and 2022-03d) with a mean time between outbursts of just $929.5\pm6.8$~days ($2.545\pm0.019$~yr), the second shortest recurrence time known for any nova. We also show that there is a blue variable source ($\langle V \rangle = 20.56\pm0.17$, $B-V\simeq0.045$), apparently coincident with the position of the nova, that exhibits a 14.3~d periodicity. Possible models of the system are proposed, but none are entirely satisfactory.

\end{abstract}


\section{Introduction} \label{sec:intro}

The nova M31N 2017-01e was first suggested to be recurrent by Tu et al. 2019\footnote{http://www.cbat.eps.harvard.edu/unconf/followups/J00441073+4154220.html}, who noted that its position closely matched that
of a nova that was seen $\sim2.5$ years later, M31N 2019-09d. Then, after another $\sim$2.5 years,
M31N 2022-03d was flagged by Zhao et al. 2022\footnote{http://www.cbat.eps.harvard.edu/unconf/followups/J00441072+4154224.html} as a likely third eruption of this nova. To firmly establish the association of these three outbursts,
\citet{2022ATel15729....1S} confirmed though registration of the outburst images that the three novae were indeed spatially coincident. They also established that M31N 2017-01e had a mean recurrence time of just $2.55\pm0.03$ years, making it the second shortest known for any recurrent nova system.

\section{Discussion}

\subsection{The 2012 eruption}

Here, we present evidence for a previously unrecognized eruption of the M31N 2017-01e system based on Pan-STARRS archival data.
The lightcurve of the source, PS1\_DR2\_11.04464+41.9061, presented in
Figure~\ref{fig:f1}, shows a clear increase in flux on 11 January 2012. A careful alignment
of the Pan-STARRS image with that of M31N 2017-01e shows that the two events are spatially
coincident (see the lower left panel of Figure~\ref{fig:f1}).

\subsection{Discovery of a Quiescent Counterpart?}

\begin{figure*}
\includegraphics[angle=0,scale=0.6]{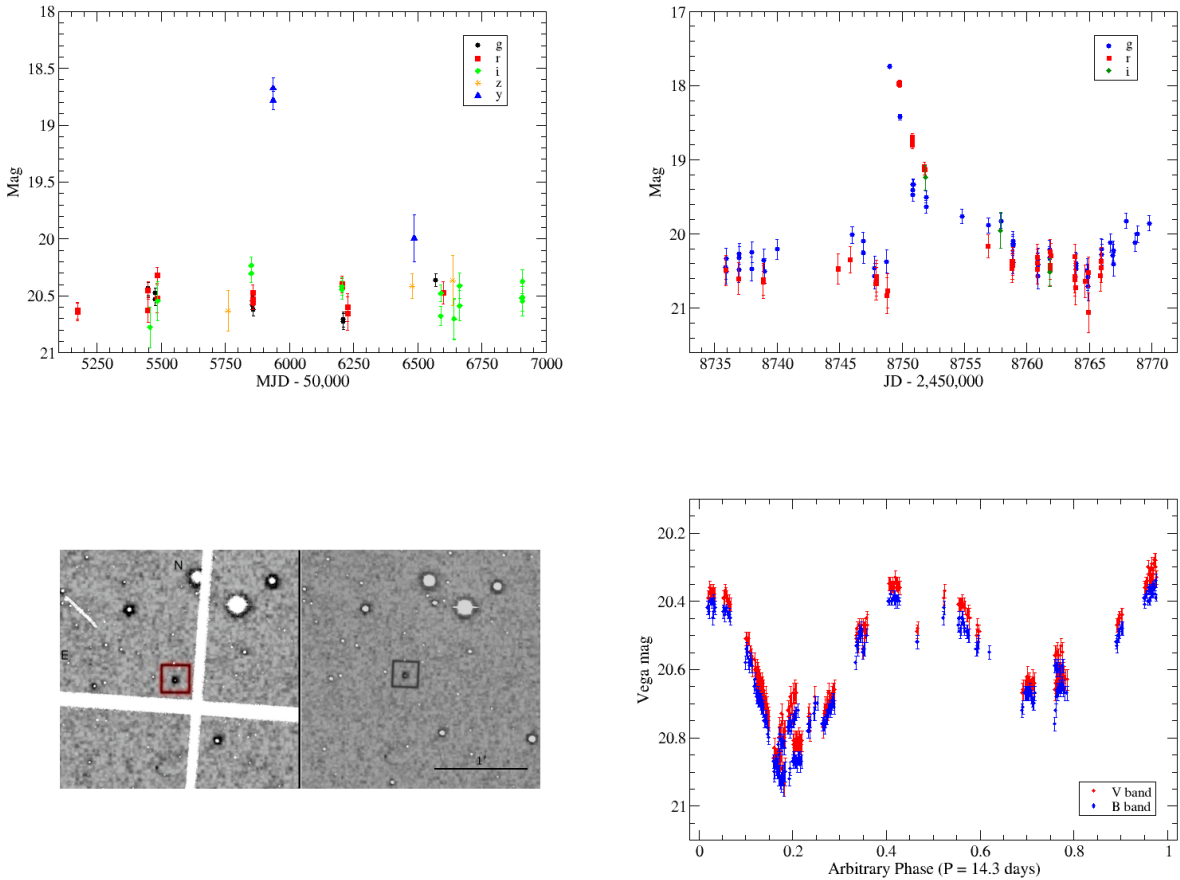}
\caption{{\it Top Left Panel:} The Pan-STARRS (PS1) lightcurve for the 11 January 2012 eruption of PS1\_DR2\_11.04464+41.9061 (which we now designate as M31N 2012-01c). {\it Top Right Panel:} The ZTF lightcurve for the 22 September 2019 eruption of M31N 2017-01e. The lightcurve shows that 
$t_2$ -- the time to fade 2 mag -- is just $\sim$6~d.
{\it Bottom Left Panel:} The positions of M31N 2012-01c and 2017-01e are compared on the left showing that they are indeed spatially coincident (white: M31N 2012-01c; Black: M31N 2017-01e), while M31N 2017-01e is compared on the right with a deep field image from \citet{2006AJ....131.2478M} showing that it is also spatially coincident with the blue, $V\sim20.4$ source, which we identify with the quiescent nova (white: M31V~J00441070+4154220; black: M31N 2017-01e).
{\it Bottom right panel:} The phased $B$ and $V$ lightcurve of M31V~J00441070+4154220 from the data from \cite{2006A&A...459..321V}. Both lightcurves exhibit an asymmetrical modulation throughout the putative 14.3~d orbital cycle of the quiescent nova.
The lightcurve data can be found in the Appendix.
}
\label{fig:f1}
\end{figure*}

Deep images of the field surrounding M31N 2017-01e from \cite{2006AJ....131.2478M} revealed that there is a faint ($V\sim20.4$) blue star, M31V~J00441070+4154220, possibly coincident with the position of the nova. More interestingly, observations by \cite{2006A&A...459..321V} had previously established that the source was variable (possibly an eclipsing binary) with a period of approximately 14.3~d (see the lightcurve in the lower right panel of Figure~\ref{fig:f1}). This variability, coupled with the spatial coincidence, suggests that the variable source and the recurrent nova are likely related. We speculate on that relationship further below.

\subsection{The Mean Recurrence Time}

With the addition of a fourth eruption from January 2012, we are now in a position to update the
mean recurrence time of the nova, and to assess its stability.
Based on the dates of the four recorded eruptions (MJD 55937.2, 57784.4, 58748.64, and 59643.59)
we are able to confirm
that the nova recurs quite regularly, with a likely eruption near the end of July of 2014 being apparently missed. Our updated mean recurrence time
is $\langle P_\mathrm{rec}\rangle=929.5\pm6.8$~d ($2.545\pm0.019$ yr), with
the full ephemeris given by: MJD$_\mathrm{erupt} = (55937.2\pm18.2) + (929.5\pm6.8)$~E\footnote{Given that the nova lies $\sim41.5'$ from the nucleus of M31 where temporal coverage of the galaxy is poor, a recurrence time of half this value (i.e., $\sim$1.25~yr), though very unlikely, may not be entirely ruled out by existing observations.}.
Based on the regularity of the eruptions
and our updated ephemeris, we expect the next eruption of M31N 2017-01e to occur on 01 October 2024 plus or minus $\sim$3 weeks.

\subsection{Possible Models}

Although a detailed model for M31N 2017-01e is beyond the scope of this {\it Research Note\/},
below we speculate on possible interpretations of the available data.
Assuming that the
optical counterpart is in fact the progenitor binary, then M31N 2017-01e is either an extremely luminous (at quiescence) recurrent nova system in M31 itself, or it is a foreground Galactic nova projected against the field of that galaxy. We briefly explore these two possibilities in turn.

At the distance of M31, $\mu_o = 24.38\pm0.05$ \citep{2001ApJ...553...47F}, a $V\sim20.4$ mag source
would have to have an absolute visual magnitude, corrected for foreground absorption
(A$_V\sim0.2$ mag), of $M_\mathrm{V}\sim-4.2$. Although unusually luminous for a nova progenitor, it is not without precedent. For example, the Galactic nova RS Oph has an absolute visual magnitude, $M_\mathrm{V}=-4.1$ \citep{2010ApJS..187..275S}. The difference, however, is that the RS~Oph has an orbital period of 453.6~d and its quiescent color ($B-V\simeq0.4$, corrected for absorption) is dominated by the red giant secondary star. The putative M31N 2017-01e progenitor star, on the other hand, is quite blue ($B-V\simeq0.045$) and exhibits a 14.3~d periodicity, which if interpreted as the orbital period, would severely limit the luminosity of the secondary star.

Could the quiescent light of M31N 2017-01e be dominated by an extremely luminous accretion disk in the system?
Even though the short recurrence time of this nova {\it requires\/} a high mass white dwarf accreting at a high rate\footnote{Spectroscopic observations of M31N 2017-01e by \citet{2017ATel10042....1W} identifying the nova as
a member of the He/N spectroscopic class \citep{1992AJ....104..725W} and showing that the Balmer lines extend to a FWZI of $\sim$10,000~km~s$^{-1}$, along with the short $t_2$ time observed in the 2019 eruption (see the top right panel of Figure\ref{fig:f1}), are consistent with an outburst arising from a nova progenitor containing a massive white dwarf.}, it appears unlikely that a disk could produce the required luminosity.
According to the models of \citet{2014ApJ...793..136K}, a recurrence time of $\sim$2.5 yr can be achieved in a system characterized by $M_\mathrm{WD} = 1.3$~M$_{\odot}$, $R_\mathrm{WD}=3\times10^8$~cm, and $\dot M = 4\times10^{-7}$~M$_{\odot}$~yr$^{-1}$. The accretion disk in such a system could produce a maximum
bolometric luminosity, $L_\mathrm{bol} \sim \frac{1}{2} (G~M_\mathrm{WD}~\dot M)/R_\mathrm{WD}$, or just $\sim$1900~L$_{\odot}$ ($M_\mathrm{bol}\sim-3.5$), which falls well short of the required absolute visual magnitude of $M_V=-4.2$ (particularly considering that the bolometric correction is likely several magnitudes). We note that such a disk might be able to contribute enough light to explain the observed brightness at quiescence, but only if the secondary star also contributes significantly to the quiescent luminosity, and this does not appear possible in a system with a 14.3~d period.

Possible models where the nova is confined to the Galaxy suffer the opposite problem: the high
white dwarf mass and accretion rate required to produce outbursts every few years would likely produce
a quiescent disk luminosity, when combined with that of the secondary star, of at least 100~L$_\odot$. A quiescent nova binary with $M_V\sim0$, then would place the putative $V\sim20.4$ mag nova progenitor at a distance of $\sim$100~kpc (the blue color makes significant extinction highly unlikely), well outside the confines of the Milky Way.

Perhaps the most promising scenario is that the orbital period of the nova is in fact much longer than assumed, possibly of order hundreds of days, and that observed 14.3~d periodicity reflects some other clock in the system, perhaps related to resonances in the accretion disk itself.
Finally, it is also remotely possible that M31N 2017-01e is a ``typical" rapidly-recurring nova in M31 and the proposed optical counterpart is just a chance positional coincidence of an unrelated binary star. However, given the extremely close spatial alignment shown in Figure~\ref{fig:f1}, we consider this possibility to be extremely unlikely.

We conclude here by noting that with the addition of the 2012 eruption, M31N 2017-01e now has had four recorded eruptions. Only M31N 1926-07c, 1963-09c, and 2008-12a with their 5, 6, and 14 observed outbursts, respectively, have been seen to erupt more frequently. As with all of these systems, it will be important to secure both photometric and spectroscopic observations of M31N 2017-01e when it is expected to erupt again in late 2024 in order to further our understanding of this fascinating recurrent nova.

\begin{acknowledgments}
We would like to thank Wenjie Zhou (周文杰) for discussion and Naoto Kojiguchi for assistance with downloading archival data used in this work.
\end{acknowledgments}

\vspace{5mm}
\facilities{Archival data from Pan-STARRS and the Zwicky Transient Facility (ZTF) were used in this work.}

\newpage

\bibliography{M31N2017-01e}{}
\bibliographystyle{aasjournal}


\appendix

\startlongtable
\begin{deluxetable}{ccccccc}
\tablenum{A1}
\tablecolumns{7}
\tablecaption{PS1 Nova Photometry for M31N 2012-01c}
\tablehead{\colhead{MJD} & \colhead{} & \colhead{} && \colhead{MJD} & \colhead{} & \colhead{}\\
\colhead{($-50,000$)} & \colhead{Filter} & \colhead{Mag} && \colhead{($-50,000$)} & \colhead{Filter} & \colhead{Mag}}
\startdata
5448.556 &$g$&$20.43\pm0.06$&&6226.327 &$r$&$20.66\pm0.15$\cr
5448.567 &$g$&$20.45\pm0.06$&&6599.225 &$r$&$20.47\pm0.10$\cr
5477.289 &$g$&$20.53\pm0.05$&&5457.543 &$i$&$20.78\pm0.18$\cr
5477.301 &$g$&$20.48\pm0.04$&&5486.274 &$i$&$20.55\pm0.17$\cr
5857.349 &$g$&$20.55\pm0.05$&&5850.468 &$i$&$20.23\pm0.08$\cr
5857.363 &$g$&$20.58\pm0.05$&&5850.481 &$i$&$20.30\pm0.08$\cr
5858.291 &$g$&$20.62\pm0.05$&&6205.437 &$i$&$20.45\pm0.08$\cr
5858.303 &$g$&$20.57\pm0.05$&&6205.449 &$i$&$20.42\pm0.08$\cr
6209.318 &$g$&$20.73\pm0.07$&&6589.450 &$i$&$20.68\pm0.09$\cr
6209.331 &$g$&$20.71\pm0.07$&&6589.462 &$i$&$20.48\pm0.07$\cr
6570.439 &$g$&$20.36\pm0.06$&&6641.284 &$i$&$20.70\pm0.18$\cr
5174.317 &$r$&$20.63\pm0.08$&&6641.296 &$i$&$20.70\pm0.18$\cr
5174.318 &$r$&$20.64\pm0.08$&&6663.227 &$i$&$20.59\pm0.13$\cr
5448.583 &$r$&$20.45\pm0.08$&&6663.240 &$i$&$20.41\pm0.11$\cr
5448.595 &$r$&$20.62\pm0.11$&&6907.435 &$i$&$20.52\pm0.15$\cr
5485.290 &$r$&$20.32\pm0.07$&&6907.446 &$i$&$20.51\pm0.12$\cr
5485.302 &$r$&$20.52\pm0.12$&&6907.458 &$i$&$20.55\pm0.13$\cr
5857.376 &$r$&$20.47\pm0.07$&&6907.469 &$i$&$20.38\pm0.11$\cr
5857.388 &$r$&$20.54\pm0.08$&&5760.585 &$z$&$20.63\pm0.18$\cr
5858.265 &$r$&$20.53\pm0.06$&&6477.598 &$z$&$20.41\pm0.11$\cr
5858.277 &$r$&$20.56\pm0.06$&&6637.214 &$z$&$20.36\pm0.22$\cr
6205.343 &$r$&$20.40\pm0.07$&&5937.201 &$y$&$18.67\pm0.09$\cr
6205.355 &$r$&$20.42\pm0.08$&&5937.213 &$y$&$18.78\pm0.08$\cr
6226.315 &$r$&$20.60\pm0.12$&&6486.625 &$y$&$20.00\pm0.21$\cr
\enddata
\end{deluxetable}

\newpage

\startlongtable
\begin{deluxetable}{ccccccccccc}
\tablenum{A2}
\tablecolumns{11}
\tablecaption{ZTF Nova Photometry for M31N 2019-09d}
\tablehead{\colhead{JD} & \colhead{} & \colhead{} && \colhead{JD} & \colhead{} & \colhead{}&& \colhead{JD} & \colhead{} & \colhead{}\\
\colhead{($-2,450,000$)} & \colhead{Filter} & \colhead{Mag} && \colhead{($-2,450,000$)} & \colhead{Filter} & \colhead{Mag}&& \colhead{($-2,450,000$)} & \colhead{Filter} & \colhead{Mag}
}
\startdata
8735.911&$g$&$20.44\pm0.16$&&8760.853&$g$&$20.41\pm0.16$&&8748.820&$r$&$20.77\pm0.24$\cr
8735.934&$g$&$20.51\pm0.17$&&8760.865&$g$&$20.34\pm0.15$&&8749.757&$r$&$17.97\pm0.03$\cr
8735.935&$g$&$20.34\pm0.15$&&8760.874&$g$&$20.56\pm0.17$&&8749.788&$r$&$17.96\pm0.03$\cr
8736.958&$g$&$20.48\pm0.17$&&8760.885&$g$&$20.42\pm0.16$&&8749.812&$r$&$17.99\pm0.03$\cr
8736.968&$g$&$20.27\pm0.14$&&8761.790&$g$&$20.22\pm0.14$&&8750.810&$r$&$18.69\pm0.05$\cr
8736.987&$g$&$20.32\pm0.15$&&8761.793&$g$&$20.32\pm0.15$&&8750.821&$r$&$18.72\pm0.05$\cr
8737.969&$g$&$20.47\pm0.16$&&8763.893&$g$&$20.40\pm0.16$&&8750.830&$r$&$18.80\pm0.05$\cr
8738.013&$g$&$20.25\pm0.14$&&8763.904&$g$&$20.47\pm0.16$&&8751.788&$r$&$19.14\pm0.07$\cr
8738.971&$g$&$20.36\pm0.15$&&8763.933&$g$&$20.42\pm0.16$&&8751.797&$r$&$19.10\pm0.07$\cr
8739.003&$g$&$20.50\pm0.17$&&8764.842&$g$&$20.49\pm0.17$&&8751.824&$r$&$19.14\pm0.07$\cr
8740.018&$g$&$20.20\pm0.13$&&8764.864&$g$&$20.58\pm0.18$&&8756.900&$r$&$20.17\pm0.16$\cr
8746.012&$g$&$20.01\pm0.12$&&8764.887&$g$&$20.71\pm0.19$&&8758.814&$r$&$20.46\pm0.20$\cr
8746.901&$g$&$20.10\pm0.12$&&8765.947&$g$&$20.20\pm0.13$&&8758.823&$r$&$20.42\pm0.19$\cr
8746.922&$g$&$20.25\pm0.14$&&8765.965&$g$&$20.28\pm0.14$&&8758.824&$r$&$20.37\pm0.18$\cr
8747.824&$g$&$20.46\pm0.16$&&8765.985&$g$&$20.20\pm0.13$&&8760.806&$r$&$20.31\pm0.18$\cr
8747.826&$g$&$20.57\pm0.17$&&8766.668&$g$&$20.12\pm0.13$&&8760.827&$r$&$20.39\pm0.19$\cr
8748.761&$g$&$20.37\pm0.15$&&8766.882&$g$&$20.28\pm0.14$&&8760.828&$r$&$20.47\pm0.20$\cr
8749.010&$g$&$17.74\pm0.03$&&8766.906&$g$&$20.23\pm0.14$&&8761.847&$r$&$20.42\pm0.19$\cr
8749.831&$g$&$18.42\pm0.04$&&8766.931&$g$&$20.41\pm0.16$&&8761.848&$r$&$20.49\pm0.20$\cr
8749.844&$g$&$18.42\pm0.04$&&8767.922&$g$&$19.82\pm0.10$&&8761.849&$r$&$20.24\pm0.17$\cr
8750.872&$g$&$19.41\pm0.07$&&8768.666&$g$&$20.11\pm0.13$&&8761.902&$r$&$20.30\pm0.17$\cr
8750.873&$g$&$19.33\pm0.07$&&8768.839&$g$&$20.00\pm0.12$&&8763.820&$r$&$20.58\pm0.21$\cr
8750.886&$g$&$19.48\pm0.08$&&8769.755&$g$&$19.85\pm0.10$&&8763.851&$r$&$20.31\pm0.17$\cr
8750.907&$g$&$19.34\pm0.07$&&8735.888&$r$&$20.49\pm0.20$&&8763.852&$r$&$20.61\pm0.22$\cr
8751.909&$g$&$19.63\pm0.09$&&8736.916&$r$&$20.60\pm0.22$&&8763.868&$r$&$20.72\pm0.23$\cr
8751.925&$g$&$19.50\pm0.08$&&8738.891&$r$&$20.61\pm0.22$&&8764.636&$r$&$20.63\pm0.22$\cr
8754.793&$g$&$19.76\pm0.10$&&8738.891&$r$&$20.65\pm0.22$&&8764.917&$r$&$20.52\pm0.21$\cr
8756.909&$g$&$19.88\pm0.11$&&8744.879&$r$&$20.47\pm0.20$&&8764.926&$r$&$21.06\pm0.27$\cr
8757.927&$g$&$19.82\pm0.10$&&8745.856&$r$&$20.35\pm0.18$&&8765.881&$r$&$20.56\pm0.21$\cr
8758.855&$g$&$20.13\pm0.13$&&8747.908&$r$&$20.67\pm0.22$&&8765.904&$r$&$20.37\pm0.18$\cr
8758.861&$g$&$20.16\pm0.13$&&8747.926&$r$&$20.67\pm0.22$&&8765.930&$r$&$20.46\pm0.20$\cr
8758.864&$g$&$20.39\pm0.15$&&8747.933&$r$&$20.57\pm0.21$&&8751.853&$i$&$19.24\pm0.17$\cr
8758.865&$g$&$20.14\pm0.13$&&8747.955&$r$&$20.61\pm0.22$&&8757.870&$i$&$19.95\pm0.24$\cr
8758.881&$g$&$20.09\pm0.12$&&8748.811&$r$&$20.83\pm0.24$&&8761.815&$i$&$20.52\pm0.19$\cr
\enddata
\end{deluxetable}

\newpage

\startlongtable
\begin{deluxetable}{ccccccccccc}
\tablenum{A3}
\tablecolumns{11}
\tablecaption{Phased M31V~J00441070+4154220 Photometry}
\tablehead{\colhead{Phase\tablenotemark{a}} & \colhead{Filter} & \colhead{Mag} && \colhead{Phase\tablenotemark{a}} & \colhead{Filter} & \colhead{Mag} && \colhead{Phase\tablenotemark{a}} & \colhead{Filter} & \colhead{Mag}
}
\startdata
0.0521 &B&$20.43\pm0.02$&&0.1812 &B&$20.94\pm0.03$&&0.4663 &V&$20.49\pm0.02$\cr
0.0539 &B&$20.42\pm0.01$&&0.1827 &B&$20.90\pm0.03$&&0.4677 &V&$20.48\pm0.02$\cr
0.0556 &B&$20.44\pm0.02$&&0.1843 &B&$20.90\pm0.03$&&0.7601 &V&$20.56\pm0.03$\cr
0.0573 &B&$20.43\pm0.01$&&0.2333 &B&$20.78\pm0.02$&&0.7622 &V&$20.55\pm0.02$\cr
0.0589 &B&$20.42\pm0.01$&&0.2349 &B&$20.76\pm0.02$&&0.7646 &V&$20.55\pm0.03$\cr
0.0606 &B&$20.43\pm0.01$&&0.2364 &B&$20.78\pm0.02$&&0.7665 &V&$20.56\pm0.02$\cr
0.0622 &B&$20.44\pm0.01$&&0.2380 &B&$20.75\pm0.02$&&0.7709 &V&$20.56\pm0.03$\cr
0.0639 &B&$20.44\pm0.01$&&0.2456 &B&$20.73\pm0.02$&&0.7726 &V&$20.53\pm0.02$\cr
0.0656 &B&$20.44\pm0.02$&&0.2471 &B&$20.73\pm0.02$&&0.7763 &V&$20.55\pm0.03$\cr
0.0673 &B&$20.45\pm0.02$&&0.2487 &B&$20.70\pm0.02$&&0.8914 &V&$20.47\pm0.02$\cr
0.1185 &B&$20.65\pm0.02$&&0.2538 &B&$20.70\pm0.02$&&0.8930 &V&$20.47\pm0.02$\cr
0.1202 &B&$20.63\pm0.02$&&0.1237 &B&$20.68\pm0.02$&&0.8954 &V&$20.46\pm0.02$\cr
0.1219 &B&$20.66\pm0.02$&&0.1256 &B&$20.68\pm0.02$&&0.8970 &V&$20.45\pm0.02$\cr
0.1237 &B&$20.68\pm0.02$&&0.1271 &B&$20.69\pm0.02$&&0.8986 &V&$20.45\pm0.02$\cr
0.1253 &B&$20.68\pm0.02$&&0.1288 &B&$20.69\pm0.02$&&0.9020 &V&$20.44\pm0.02$\cr
0.1269 &B&$20.67\pm0.02$&&0.1304 &B&$20.69\pm0.02$&&0.9036 &V&$20.44\pm0.02$\cr
0.1285 &B&$20.68\pm0.02$&&0.1320 &B&$20.70\pm0.02$&&0.9616 &V&$20.36\pm0.02$\cr
0.1302 &B&$20.69\pm0.02$&&0.1336 &B&$20.71\pm0.02$&&0.9632 &V&$20.34\pm0.02$\cr
0.1319 &B&$20.69\pm0.02$&&0.1352 &B&$20.70\pm0.02$&&0.9648 &V&$20.33\pm0.02$\cr
0.1335 &B&$20.69\pm0.02$&&0.1368 &B&$20.72\pm0.02$&&0.9665 &V&$20.34\pm0.02$\cr
0.1352 &B&$20.71\pm0.02$&&0.1383 &B&$20.72\pm0.02$&&0.9681 &V&$20.34\pm0.02$\cr
0.1368 &B&$20.71\pm0.02$&&0.1399 &B&$20.75\pm0.02$&&0.9697 &V&$20.34\pm0.02$\cr
0.1419 &B&$20.75\pm0.02$&&0.1415 &B&$20.75\pm0.02$&&0.9743 &V&$20.34\pm0.02$\cr
0.1897 &B&$20.77\pm0.02$&&0.1431 &B&$20.74\pm0.02$&&0.9764 &V&$20.33\pm0.02$\cr
0.1917 &B&$20.77\pm0.02$&&0.1446 &B&$20.75\pm0.02$&&0.0311 &V&$20.39\pm0.02$\cr
0.1933 &B&$20.76\pm0.02$&&0.1462 &B&$20.77\pm0.02$&&0.0327 &V&$20.38\pm0.02$\cr
0.1950 &B&$20.78\pm0.02$&&0.1478 &B&$20.79\pm0.02$&&0.1007 &V&$20.51\pm0.02$\cr
0.1966 &B&$20.75\pm0.02$&&0.1494 &B&$20.80\pm0.02$&&0.1024 &V&$20.51\pm0.02$\cr
0.1983 &B&$20.77\pm0.02$&&0.1933 &B&$20.92\pm0.02$&&0.1040 &V&$20.53\pm0.02$\cr
0.2007 &B&$20.75\pm0.02$&&0.1953 &B&$20.89\pm0.02$&&0.1057 &V&$20.51\pm0.02$\cr
0.2024 &B&$20.75\pm0.02$&&0.2006 &B&$20.87\pm0.02$&&0.1073 &V&$20.51\pm0.02$\cr
0.2041 &B&$20.74\pm0.02$&&0.2021 &B&$20.86\pm0.02$&&0.1089 &V&$20.55\pm0.02$\cr
0.2057 &B&$20.73\pm0.02$&&0.2037 &B&$20.86\pm0.02$&&0.1105 &V&$20.55\pm0.02$\cr
0.2075 &B&$20.73\pm0.02$&&0.2052 &B&$20.87\pm0.02$&&0.1122 &V&$20.56\pm0.02$\cr
0.2123 &B&$20.72\pm0.02$&&0.2068 &B&$20.87\pm0.02$&&0.1138 &V&$20.59\pm0.03$\cr
0.5220 &B&$20.45\pm0.02$&&0.2084 &B&$20.87\pm0.02$&&0.1708 &V&$20.80\pm0.03$\cr
0.5235 &B&$20.42\pm0.02$&&0.2099 &B&$20.88\pm0.02$&&0.1724 &V&$20.77\pm0.03$\cr
0.5921 &B&$20.54\pm0.02$&&0.2116 &B&$20.86\pm0.02$&&0.1740 &V&$20.79\pm0.03$\cr
0.5935 &B&$20.54\pm0.02$&&0.2132 &B&$20.86\pm0.02$&&0.1756 &V&$20.78\pm0.03$\cr
0.5950 &B&$20.53\pm0.02$&&0.2147 &B&$20.86\pm0.02$&&0.1772 &V&$20.73\pm0.03$\cr
0.5965 &B&$20.52\pm0.02$&&0.2162 &B&$20.88\pm0.02$&&0.1789 &V&$20.76\pm0.03$\cr
0.5535 &B&$20.47\pm0.02$&&0.2179 &B&$20.86\pm0.02$&&0.1805 &V&$20.80\pm0.03$\cr
0.5559 &B&$20.45\pm0.02$&&0.2194 &B&$20.87\pm0.02$&&0.1821 &V&$20.78\pm0.03$\cr
0.5576 &B&$20.49\pm0.02$&&0.2635 &B&$20.76\pm0.02$&&0.1838 &V&$20.80\pm0.03$\cr
0.5592 &B&$20.48\pm0.02$&&0.2659 &B&$20.76\pm0.02$&&0.1846 &V&$20.78\pm0.03$\cr
0.5608 &B&$20.45\pm0.02$&&0.2674 &B&$20.76\pm0.02$&&0.9507 &V&$20.36\pm0.02$\cr
0.5624 &B&$20.45\pm0.02$&&0.2690 &B&$20.77\pm0.02$&&0.9523 &V&$20.36\pm0.02$\cr
0.5651 &B&$20.47\pm0.02$&&0.2705 &B&$20.75\pm0.02$&&0.9538 &V&$20.35\pm0.02$\cr
0.5667 &B&$20.48\pm0.02$&&0.2721 &B&$20.74\pm0.02$&&0.9554 &V&$20.35\pm0.02$\cr
0.5713 &B&$20.49\pm0.02$&&0.2736 &B&$20.74\pm0.02$&&0.9570 &V&$20.34\pm0.02$\cr
0.5729 &B&$20.49\pm0.02$&&0.2752 &B&$20.72\pm0.02$&&0.9585 &V&$20.30\pm0.02$\cr
0.5744 &B&$20.50\pm0.02$&&0.2767 &B&$20.72\pm0.02$&&0.9601 &V&$20.32\pm0.02$\cr
0.5761 &B&$20.50\pm0.02$&&0.2783 &B&$20.72\pm0.02$&&0.9616 &V&$20.32\pm0.02$\cr
0.6195 &B&$20.55\pm0.02$&&0.2798 &B&$20.71\pm0.02$&&0.9632 &V&$20.32\pm0.02$\cr
0.6891 &B&$20.72\pm0.02$&&0.2814 &B&$20.72\pm0.02$&&0.9647 &V&$20.31\pm0.02$\cr
0.6912 &B&$20.71\pm0.02$&&0.2831 &B&$20.71\pm0.02$&&0.9663 &V&$20.29\pm0.02$\cr
0.6957 &B&$20.67\pm0.02$&&0.2846 &B&$20.70\pm0.02$&&0.9678 &V&$20.32\pm0.02$\cr
0.6973 &B&$20.67\pm0.02$&&0.2862 &B&$20.69\pm0.02$&&0.9694 &V&$20.32\pm0.02$\cr
0.6989 &B&$20.67\pm0.02$&&0.2877 &B&$20.69\pm0.02$&&0.9709 &V&$20.29\pm0.02$\cr
0.7005 &B&$20.67\pm0.02$&&0.2894 &B&$20.71\pm0.02$&&0.9725 &V&$20.28\pm0.02$\cr
0.7020 &B&$20.67\pm0.02$&&0.3348 &B&$20.58\pm0.02$&&0.9741 &V&$20.28\pm0.02$\cr
0.7036 &B&$20.67\pm0.02$&&0.3364 &B&$20.53\pm0.02$&&0.9749 &V&$20.30\pm0.02$\cr
0.7067 &B&$20.66\pm0.02$&&0.3379 &B&$20.52\pm0.02$&&0.0201 &V&$20.39\pm0.02$\cr
0.7083 &B&$20.67\pm0.02$&&0.3395 &B&$20.54\pm0.02$&&0.0217 &V&$20.39\pm0.02$\cr
0.7099 &B&$20.68\pm0.02$&&0.3410 &B&$20.51\pm0.02$&&0.0232 &V&$20.36\pm0.02$\cr
0.7115 &B&$20.69\pm0.02$&&0.3425 &B&$20.49\pm0.02$&&0.0248 &V&$20.37\pm0.02$\cr
0.7130 &B&$20.70\pm0.02$&&0.3441 &B&$20.50\pm0.02$&&0.0264 &V&$20.38\pm0.02$\cr
0.7146 &B&$20.70\pm0.02$&&0.3456 &B&$20.50\pm0.02$&&0.0279 &V&$20.37\pm0.02$\cr
0.7162 &B&$20.67\pm0.02$&&0.3472 &B&$20.48\pm0.02$&&0.0295 &V&$20.38\pm0.02$\cr
0.7592 &B&$20.76\pm0.02$&&0.3487 &B&$20.50\pm0.02$&&0.0310 &V&$20.37\pm0.02$\cr
0.7612 &B&$20.72\pm0.02$&&0.3503 &B&$20.55\pm0.02$&&0.0326 &V&$20.38\pm0.02$\cr
0.7627 &B&$20.68\pm0.02$&&0.3518 &B&$20.54\pm0.02$&&0.0341 &V&$20.38\pm0.02$\cr
0.7643 &B&$20.66\pm0.02$&&0.3534 &B&$20.54\pm0.02$&&0.1598 &V&$20.86\pm0.03$\cr
0.7658 &B&$20.67\pm0.02$&&0.3580 &B&$20.50\pm0.02$&&0.1614 &V&$20.83\pm0.03$\cr
0.7675 &B&$20.66\pm0.02$&&0.4046 &B&$20.40\pm0.01$&&0.1630 &V&$20.87\pm0.03$\cr
0.7691 &B&$20.65\pm0.02$&&0.4061 &B&$20.40\pm0.01$&&0.1645 &V&$20.86\pm0.03$\cr
0.7706 &B&$20.66\pm0.02$&&0.4076 &B&$20.40\pm0.01$&&0.1661 &V&$20.86\pm0.03$\cr
0.7722 &B&$20.65\pm0.02$&&0.4092 &B&$20.38\pm0.01$&&0.1676 &V&$20.87\pm0.04$\cr
0.7738 &B&$20.65\pm0.02$&&0.4107 &B&$20.40\pm0.01$&&0.1692 &V&$20.84\pm0.03$\cr
0.7753 &B&$20.64\pm0.02$&&0.4123 &B&$20.40\pm0.01$&&0.1707 &V&$20.85\pm0.03$\cr
0.7769 &B&$20.66\pm0.02$&&0.4138 &B&$20.39\pm0.01$&&0.1724 &V&$20.87\pm0.03$\cr
0.7785 &B&$20.65\pm0.02$&&0.4154 &B&$20.39\pm0.01$&&0.1740 &V&$20.87\pm0.03$\cr
0.7813 &B&$20.67\pm0.02$&&0.4169 &B&$20.39\pm0.02$&&0.1755 &V&$20.84\pm0.03$\cr
0.7858 &B&$20.67\pm0.02$&&0.4184 &B&$20.39\pm0.02$&&0.1771 &V&$20.89\pm0.04$\cr
0.4656 &B&$20.52\pm0.02$&&0.4200 &B&$20.40\pm0.02$&&0.1804 &V&$20.91\pm0.04$\cr
0.4670 &B&$20.52\pm0.02$&&0.4215 &B&$20.39\pm0.01$&&0.1819 &V&$20.93\pm0.04$\cr
0.7590 &B&$20.59\pm0.02$&&0.4231 &B&$20.39\pm0.01$&&0.1835 &V&$20.90\pm0.04$\cr
0.7614 &B&$20.59\pm0.02$&&0.4247 &B&$20.40\pm0.02$&&0.1850 &V&$20.90\pm0.04$\cr
0.7636 &B&$20.60\pm0.02$&&0.4262 &B&$20.39\pm0.01$&&0.2341 &V&$20.74\pm0.03$\cr
0.7654 &B&$20.59\pm0.02$&&0.4277 &B&$20.40\pm0.01$&&0.2356 &V&$20.73\pm0.03$\cr
0.7696 &B&$20.60\pm0.02$&&0.0529 &V&$20.39\pm0.02$&&0.2372 &V&$20.78\pm0.03$\cr
0.7717 &B&$20.61\pm0.02$&&0.0547 &V&$20.39\pm0.02$&&0.2479 &V&$20.68\pm0.03$\cr
0.7749 &B&$20.59\pm0.02$&&0.0563 &V&$20.37\pm0.02$&&0.1245 &V&$20.66\pm0.03$\cr
0.7773 &B&$20.59\pm0.02$&&0.0581 &V&$20.38\pm0.02$&&0.1264 &V&$20.65\pm0.03$\cr
0.8906 &B&$20.52\pm0.02$&&0.0597 &V&$20.39\pm0.02$&&0.1279 &V&$20.65\pm0.03$\cr
0.8922 &B&$20.53\pm0.02$&&0.0614 &V&$20.40\pm0.02$&&0.1296 &V&$20.63\pm0.03$\cr
0.8938 &B&$20.50\pm0.02$&&0.0630 &V&$20.41\pm0.02$&&0.1312 &V&$20.65\pm0.03$\cr
0.8962 &B&$20.50\pm0.02$&&0.0647 &V&$20.40\pm0.02$&&0.1328 &V&$20.66\pm0.03$\cr
0.8978 &B&$20.49\pm0.02$&&0.0664 &V&$20.42\pm0.02$&&0.1344 &V&$20.67\pm0.03$\cr
0.9012 &B&$20.48\pm0.02$&&0.0681 &V&$20.41\pm0.02$&&0.1360 &V&$20.65\pm0.03$\cr
0.9028 &B&$20.48\pm0.02$&&0.1193 &V&$20.57\pm0.02$&&0.1376 &V&$20.67\pm0.03$\cr
0.9044 &B&$20.48\pm0.02$&&0.1210 &V&$20.60\pm0.03$&&0.1391 &V&$20.69\pm0.03$\cr
0.9608 &B&$20.40\pm0.01$&&0.1227 &V&$20.62\pm0.03$&&0.1407 &V&$20.69\pm0.03$\cr
0.9624 &B&$20.40\pm0.01$&&0.1244 &V&$20.60\pm0.03$&&0.1423 &V&$20.71\pm0.03$\cr
0.9641 &B&$20.39\pm0.01$&&0.1261 &V&$20.61\pm0.03$&&0.1439 &V&$20.70\pm0.03$\cr
0.9657 &B&$20.38\pm0.01$&&0.1277 &V&$20.62\pm0.03$&&0.1454 &V&$20.71\pm0.03$\cr
0.9673 &B&$20.38\pm0.01$&&0.1293 &V&$20.64\pm0.03$&&0.1470 &V&$20.73\pm0.03$\cr
0.9689 &B&$20.38\pm0.01$&&0.1309 &V&$20.64\pm0.03$&&0.1486 &V&$20.74\pm0.03$\cr
0.9735 &B&$20.38\pm0.01$&&0.1327 &V&$20.65\pm0.03$&&0.1502 &V&$20.75\pm0.03$\cr
0.9757 &B&$20.39\pm0.01$&&0.1343 &V&$20.64\pm0.03$&&0.1941 &V&$20.77\pm0.03$\cr
0.0303 &B&$20.45\pm0.02$&&0.1360 &V&$20.66\pm0.03$&&0.2014 &V&$20.82\pm0.03$\cr
0.0319 &B&$20.43\pm0.02$&&0.1376 &V&$20.66\pm0.03$&&0.2029 &V&$20.81\pm0.03$\cr
0.0336 &B&$20.41\pm0.01$&&0.1427 &V&$20.68\pm0.03$&&0.2045 &V&$20.83\pm0.03$\cr
0.0999 &B&$20.58\pm0.02$&&0.1905 &V&$20.72\pm0.03$&&0.2060 &V&$20.81\pm0.03$\cr
0.1016 &B&$20.54\pm0.02$&&0.1925 &V&$20.74\pm0.03$&&0.2076 &V&$20.81\pm0.03$\cr
0.1032 &B&$20.55\pm0.02$&&0.1941 &V&$20.71\pm0.03$&&0.2092 &V&$20.81\pm0.03$\cr
0.1049 &B&$20.55\pm0.02$&&0.1958 &V&$20.68\pm0.03$&&0.2107 &V&$20.83\pm0.03$\cr
0.1065 &B&$20.58\pm0.02$&&0.1974 &V&$20.70\pm0.03$&&0.2124 &V&$20.82\pm0.03$\cr
0.1081 &B&$20.59\pm0.02$&&0.1991 &V&$20.73\pm0.03$&&0.2139 &V&$20.80\pm0.03$\cr
0.1097 &B&$20.57\pm0.02$&&0.2015 &V&$20.67\pm0.03$&&0.2155 &V&$20.81\pm0.03$\cr
0.1114 &B&$20.57\pm0.02$&&0.2032 &V&$20.70\pm0.03$&&0.2170 &V&$20.82\pm0.03$\cr
0.1130 &B&$20.59\pm0.02$&&0.2049 &V&$20.66\pm0.03$&&0.2186 &V&$20.82\pm0.03$\cr
0.1700 &B&$20.82\pm0.02$&&0.2065 &V&$20.67\pm0.03$&&0.2202 &V&$20.81\pm0.03$\cr
0.1716 &B&$20.85\pm0.02$&&0.2083 &V&$20.67\pm0.03$&&0.2651 &V&$20.77\pm0.03$\cr
0.1732 &B&$20.79\pm0.02$&&0.5227 &V&$20.39\pm0.02$&&0.2666 &V&$20.75\pm0.03$\cr
0.1748 &B&$20.81\pm0.02$&&0.5242 &V&$20.37\pm0.02$&&0.2682 &V&$20.73\pm0.03$\cr
0.1765 &B&$20.82\pm0.02$&&0.5928 &V&$20.50\pm0.02$&&0.2697 &V&$20.71\pm0.03$\cr
0.1781 &B&$20.81\pm0.02$&&0.5943 &V&$20.47\pm0.02$&&0.2713 &V&$20.73\pm0.03$\cr
0.1797 &B&$20.83\pm0.02$&&0.5958 &V&$20.50\pm0.02$&&0.2729 &V&$20.71\pm0.03$\cr
0.1814 &B&$20.81\pm0.02$&&0.5972 &V&$20.49\pm0.02$&&0.2744 &V&$20.69\pm0.03$\cr
0.1830 &B&$20.83\pm0.02$&&0.5543 &V&$20.41\pm0.02$&&0.2759 &V&$20.72\pm0.03$\cr
0.9500 &B&$20.39\pm0.01$&&0.5567 &V&$20.41\pm0.02$&&0.2775 &V&$20.67\pm0.03$\cr
0.9515 &B&$20.41\pm0.01$&&0.5583 &V&$20.41\pm0.02$&&0.2791 &V&$20.69\pm0.03$\cr
0.9531 &B&$20.39\pm0.01$&&0.5599 &V&$20.41\pm0.02$&&0.2806 &V&$20.70\pm0.03$\cr
0.9546 &B&$20.38\pm0.01$&&0.5616 &V&$20.40\pm0.02$&&0.2822 &V&$20.70\pm0.03$\cr
0.9562 &B&$20.38\pm0.01$&&0.5632 &V&$20.41\pm0.02$&&0.2839 &V&$20.68\pm0.03$\cr
0.9577 &B&$20.40\pm0.01$&&0.5659 &V&$20.41\pm0.02$&&0.2854 &V&$20.67\pm0.03$\cr
0.9593 &B&$20.37\pm0.01$&&0.5675 &V&$20.42\pm0.02$&&0.2869 &V&$20.67\pm0.03$\cr
0.9608 &B&$20.36\pm0.01$&&0.5721 &V&$20.44\pm0.02$&&0.2885 &V&$20.67\pm0.03$\cr
0.9624 &B&$20.37\pm0.01$&&0.5736 &V&$20.44\pm0.02$&&0.2901 &V&$20.66\pm0.03$\cr
0.9640 &B&$20.36\pm0.01$&&0.5752 &V&$20.43\pm0.02$&&0.3356 &V&$20.50\pm0.02$\cr
0.9655 &B&$20.37\pm0.01$&&0.5768 &V&$20.45\pm0.03$&&0.3371 &V&$20.51\pm0.02$\cr
0.9671 &B&$20.36\pm0.01$&&0.6899 &V&$20.67\pm0.03$&&0.3387 &V&$20.48\pm0.02$\cr
0.9686 &B&$20.36\pm0.01$&&0.6920 &V&$20.66\pm0.03$&&0.3402 &V&$20.52\pm0.03$\cr
0.9702 &B&$20.35\pm0.01$&&0.6965 &V&$20.66\pm0.03$&&0.3418 &V&$20.47\pm0.02$\cr
0.9717 &B&$20.36\pm0.01$&&0.6981 &V&$20.65\pm0.03$&&0.3433 &V&$20.50\pm0.02$\cr
0.9733 &B&$20.36\pm0.02$&&0.6997 &V&$20.65\pm0.03$&&0.3448 &V&$20.48\pm0.02$\cr
0.9756 &B&$20.34\pm0.02$&&0.7013 &V&$20.62\pm0.03$&&0.3464 &V&$20.50\pm0.02$\cr
0.0194 &B&$20.42\pm0.02$&&0.7028 &V&$20.64\pm0.03$&&0.3480 &V&$20.47\pm0.02$\cr
0.0209 &B&$20.42\pm0.02$&&0.7044 &V&$20.65\pm0.03$&&0.3495 &V&$20.54\pm0.03$\cr
0.0225 &B&$20.43\pm0.02$&&0.7059 &V&$20.64\pm0.03$&&0.3511 &V&$20.47\pm0.02$\cr
0.0240 &B&$20.40\pm0.01$&&0.7075 &V&$20.66\pm0.03$&&0.3526 &V&$20.49\pm0.03$\cr
0.0256 &B&$20.40\pm0.01$&&0.7091 &V&$20.66\pm0.03$&&0.3542 &V&$20.49\pm0.03$\cr
0.0271 &B&$20.40\pm0.01$&&0.7107 &V&$20.65\pm0.03$&&0.3557 &V&$20.48\pm0.03$\cr
0.0287 &B&$20.41\pm0.01$&&0.7122 &V&$20.63\pm0.03$&&0.3588 &V&$20.45\pm0.02$\cr
0.0302 &B&$20.40\pm0.01$&&0.7138 &V&$20.65\pm0.03$&&0.3596 &V&$20.47\pm0.03$\cr
0.0318 &B&$20.40\pm0.02$&&0.7154 &V&$20.64\pm0.03$&&0.4053 &V&$20.35\pm0.02$\cr
0.0334 &B&$20.41\pm0.02$&&0.7170 &V&$20.64\pm0.03$&&0.4069 &V&$20.35\pm0.02$\cr
0.0349 &B&$20.42\pm0.02$&&0.7600 &V&$20.66\pm0.03$&&0.4084 &V&$20.35\pm0.02$\cr
0.1591 &B&$20.87\pm0.02$&&0.7619 &V&$20.64\pm0.03$&&0.4100 &V&$20.36\pm0.02$\cr
0.1606 &B&$20.90\pm0.03$&&0.7635 &V&$20.64\pm0.03$&&0.4115 &V&$20.36\pm0.02$\cr
0.1622 &B&$20.89\pm0.03$&&0.7651 &V&$20.64\pm0.03$&&0.4131 &V&$20.35\pm0.02$\cr
0.1637 &B&$20.87\pm0.02$&&0.7666 &V&$20.61\pm0.03$&&0.4146 &V&$20.37\pm0.02$\cr
0.1653 &B&$20.88\pm0.03$&&0.7683 &V&$20.64\pm0.03$&&0.4161 &V&$20.36\pm0.02$\cr
0.1668 &B&$20.89\pm0.03$&&0.7698 &V&$20.61\pm0.03$&&0.4177 &V&$20.36\pm0.02$\cr
0.1684 &B&$20.89\pm0.03$&&0.7714 &V&$20.63\pm0.03$&&0.4192 &V&$20.33\pm0.02$\cr
0.1700 &B&$20.91\pm0.03$&&0.7730 &V&$20.61\pm0.03$&&0.4208 &V&$20.35\pm0.02$\cr
0.1716 &B&$20.91\pm0.03$&&0.7746 &V&$20.62\pm0.03$&&0.4223 &V&$20.35\pm0.02$\cr
0.1732 &B&$20.92\pm0.02$&&0.7761 &V&$20.62\pm0.03$&&0.4239 &V&$20.35\pm0.02$\cr
0.1748 &B&$20.93\pm0.02$&&0.7777 &V&$20.61\pm0.03$&&0.4254 &V&$20.37\pm0.02$\cr
0.1763 &B&$20.93\pm0.03$&&0.7792 &V&$20.62\pm0.03$&&0.4270 &V&$20.35\pm0.02$\cr
0.1779 &B&$20.91\pm0.03$&&0.7821 &V&$20.63\pm0.03$&&0.4285 &V&$20.36\pm0.02$\cr
0.1796 &B&$20.91\pm0.03$&&0.7866 &V&$20.63\pm0.03$&&0.4293 &V&$20.36\pm0.02$\cr
\enddata
\tablenotetext{a}{Arbitrary phase based on $\mathrm{HJD}_0=2,451,431.7000$ and $P=14.2999$~d.}
\end{deluxetable}

\end{CJK*}
\end{document}